# Comprehensive Performance Evaluation of LID Practices for the Sponge City Construction: A Case Study in Guangxi, China

Qian Li[a, b, c, #], Feng Wang[a, b, c, #], Yang Yu[a, b], Zhengce Huang[d], Mantao Li[d], Yuntao Guan[a, b, c, *]

[a] Graduate School at Shenzhen, Tsinghua University, Shenzhen 518055, P.R. China

[b] Guangdong Provincial Engineering Technology Research Center for Urban Water Cycle and Water Environment Safety, Shenzhen 518055, P.R. China

[c] State Environmental Protection Key Laboratory of Microorganism Application and Risk Control, School of Environment, Tsinghua University, Beijing 10084, P.R. China

[d] Municipal Design Institute in Hualan group, Nanning 530011, P.R. China

\# These authors contributed equally to this work

* Corresponding author: Email: guanyt@tsinghua.edu.cn, Tel.: +86-755-26036702, Fax: +86-755-26036702, Address: Graduate School at Shenzhen, Tsinghua University, Shenzhen 518055, P.R. China

**ABSTRACT** Sponge city construction is a new concept of urban stormwater management, which can effectively relieve urban flooding, reduce non-point source pollution, and promote the usage of rainwater resources, often including the application of Low Impact Development (LID) techniques. Although 30 cities in China have been chosen to implement sponge city construction, there is a lack of a quantitative evaluation method to evaluate the environmental, economic, and social benefits of LID practices. This paper develops a comprehensive evaluation system to quantify the benefits of different combinations of LID units using the Storm Water Management Model (SWMM) and the Analytical Hierarchy Process (AHP) method. The performance of five LID design scenarios with different locations and sizes of the bio-retention facility, the grassed swale, the sunken green space, the permeable pavement, and the storage tank were analyzed for a sports center project in Guangxi, China. Results indicated that the green scenario that contains 34.5% of bio-retention facilities and 46.0% of sunken green spaces had the best comprehensive performance regarding meeting the requirements of 75% annual total runoff reduction and the attainment of good operation performance, rainwater utilization, landscape promotion, and ecological service functions, mainly because they are micro-scale and decentralized facilities that can manage stormwater at the source through the natural process. The optimal scenario was adopted to construct the project, and the proposed evaluation system can also be applied to optimal selection and performance effect evaluation of LID practices in other sponge city projects.

**KEYWORDS:** sponge city; performance evaluation; LID; SWMM; AHP

# 1. Introduction

Large-scale urbanization in recent years has led to a rapid increase in impermeable surface areas. These areas have changed the natural hydrologic cycle (Jacobson, 2011; Zhang et al., 2014), and resulted in severe flooding, runoff pollution, water environment deterioration, and ecological damage (Paule-Mercado et al., 2017; Zhou, 2014). Traditional stormwater management methods are not capable of completely meeting the goals of sustainable urban development (Qin et al., 2013). On the other hand, sponge city construction is a new concept of urban stormwater management that provides a natural and low impact way to manage stormwater (Ahiablame et al., 2012; Kong et al., 2017).

The basic concept of a sponge city does not only refer to Low Impact Development (LID). It includes LID at the source, the stormwater drainage pipe system at the midway, and excessive stormwater drainage system at the terminal such as deep tunnel drainage systems and natural water bodies. The combination of LID techniques and systems could enhance the resilience of cities to cope with environmental risks from storm events of various recurrence periods (Casal-Campos et al., 2015; Xie et al., 2017). However, the principle of conventional stormwater drainage systems mainly focuses on "rapid-draining" of rainwater to downstream rather than retain and reuse it, which is contrary to the idea of constructing a city as a "resilient sponge" and thus is considered unsustainable (Barbosa et al., 2012; Chen et al., 2016; Qin et al., 2013). Instead, the LID techniques are considered ideal measures to improve the urban resilience, which manage stormwater at the source through the natural processes of infiltration, detention, storage, and purification (Dong et al., 2017; Hunt et al., 2010; Jia et al., 2017). Moreover, since the LID techniques usually are micro-scale and decentralized facilities, they are more feasible to be applied to highly urbanized areas in comparison with the upgrade of large-scale underground drainage systems (Gregoire and Clausen, 2011). Therefore, the concept and performance of sponge city construction can be notably embodied by LID practices.

In general, the aim of sponge city construction is to maintain as much as possible an unchanged regional hydrologic cycle in the process of rapid city-development. Also, sponge city construction takes into account the water resources, the water environment, the water security, the water economy, and the water cultural aspects (Zuo, 2016). This type of design that involves LID techniques can not only provide environmental benefits but also accelerate economic and social sustainable development (Demuzere et al., 2014). However, a majority of the existing studies mainly focus on the environmental benefits of LID practices, such as the control of runoff quantity and quality (Baek et al., 2015; Davis et al., 2012b; Liu et al., 2016). A limited number of studies have investigated the comprehensive benefits of LID facilities and optimization of LID combinations. For instance, Jia et al. (2015) attempted to simulate the efficiency of stormwater control using LID facilities with bio-retention, green roofs, grassed swales, rainwater storage tanks, and permeable pavements. They applied a decision support tool to evaluate the benefits of these LID facilities in view of water quality and quantity control and implementation cost (i.e., construction cost and maintenance cost) (Jia et al., 2015). Similarly, Chui et al. (2016) investigated the optimal LID scheme focusing on the lowest cost and the reduction of at least 20% of peak runoff. The LID facility unit cost to reduce peak runoff was calculated for Hong Kong and Seattle (Chui et al., 2016). In contrast, J. Li et al. (2017) suggested a preferred order for each LID facility consisting of a bio-retention facility, a green roof, a storage tank, and a sunken green space. They then evaluated the performance of LID combinations consisting of two or three components (J. Li et al., 2017).

Although the environmental benefits and implementation cost of LID facilities are essential factors when choosing an optimal LID scheme, the multiple benefits provided by these facilities, such as operation performance and social benefits (e.g., the landscape function and ecological service functions) (Pauleit et al., 2011), are equally important in decision making for sponge city construction (Visitacion et al., 2009). Additionally, existing evaluation systems do not consider a detailed optimization of sizes and locations of LID facilities, even though this is a priority in LID design schemes (Gilroy and McCuen, 2009).

This study proposes a comprehensive evaluation system, with an aim to quantify the performance of LID practices from the perspective of providing environmental, economic, and social benefits as well as provide references for optimization of types, locations, and sizes of LID facilities on the basis of scorings, comparisons, and the screening of various LID combination scenarios. The Storm Water Management Model (SWMM) is applied to obtain LID environmental indicator values, mainly because it is an open source software and has been widely applied to simulate water volumes, pollutant loads and LID practices, and is comparable to and compatible with its counterparts, e.g., InfoWorks and MIKE URBAN regarding one-dimensional modeling (Bosley II, 2008; Jacobson, 2011; Koudelak and West, 2008). In addition to the evaluation of the environmental benefits, the economic and social benefits are evaluated using the Analytical Hierarchy Process (AHP) method which takes into account multiple indicators (Rahmati et al., 2016; Saaty, 2008). Finally, a LID scenario is developed that accomplishes the goals of a sponge city that not only mimics predevelopment hydrology but also provides significant economic and social benefits. Simultaneously, various size ranges of LID facilities are suggested for the design of LID schemes for use in different site conditions.

## 2. Methodology

### 2.1. Study area

The proposed evaluation system was applied to the Guangxi Sports Center (Fig. 1), which is located in Nanning, China. Nanning is the capital city of the Guangxi Province and serves as a regional commercial and economic center and has a relatively high average annual rainfall volume of 1304.2 mm. Therefore, this site was selected as a pilot city for the implementation of sponge city construction with the primary aim to reduce at least 75% and 80% of the annual rainfall runoff for reconstruction and new projects, respectively. Several sponge city projects have been built, including the sports center, which is the largest sports stadium in the city with a total area of 64.61 ha that includes 22.98 ha or 35.6% green land, 12.56 ha or 19.4% roof areas, and 29.07 ha or 45.0% roads. The stormwater pipe networks for this sports center collect and discharge rainfall runoff into municipal pipe networks through six links with different diameters, notated as A to F in Fig. 2. The study area was then divided into six sub-catchments based on the current land use and stormwater pipe networks, where each sub-catchment corresponds to an outfall.

Insert Fig. 1

**Fig. 1.** Land uses of the Guangxi sports center.

Insert Fig. 2

**Fig. 2.** Diagram of the stormwater pipe networks for the sports center.

**2.2. Development of the LID scenarios**

The primary purpose of sponge city construction is to reduce runoff, delay peak flows, and mitigate non-point source pollution. During the initial stages of LID scenario development, the Annual Total Runoff Control Rate (ATRCR) needs to be calculated. Since this project is classified as a reconstruction project, the ATRCR should not be less than 75%. Hence, a 75% ATRCR was set to meet the minimum requirement for the whole study area, although the runoff control rate may more or less than 75% for different sub-catchments. To simplify the total runoff control process in the development of LID scenarios and make it feasible in engineering practices, normally the ATRCR corresponds to a certain rainfall depth from a statistical point of view. Based on the statistics of historical rainfall events, the relationship between the ATRCR and rainfall depth in Nanning City was examined (Fig. 3), from which a rainfall depth of 26 mm was used for the case of 75% ATRCR.

Insert Fig. 3

**Fig. 3.** The relationship between rainfall depth and ATRCR in Nanning City.

Based on the current status investigation, it was found that there have five stormwater storage tanks built in the sports center that can hold a total effective volume of 1,108 m$^3$ (Fig. 1). Also, the existing sunken green space is capable of controlling 571 m$^3$ of stormwater. Moreover, there is a large infiltration pond in the southeast of the sports center that can accept rainfall runoff and then infiltrate and purify rainwater, which can control around 50 m$^3$ of rainfall runoff. Overall, the existing stormwater management facilities can control 1,729 m$^3$ of runoff, with a corresponding rainfall depth of 4.5 mm, according to the runoff calculation obtained from Equation (1):

$$W = 10 \cdot \Psi_c \cdot h \cdot F \qquad (1)$$

where W is the rainfall runoff volume (m$^3$), $\Psi_c$ is the comprehensive runoff coefficient (0.59 in this case, as shown in Table 1), h is the rainfall depth (mm), and F is the catchment area (ha) (64.61 ha in this case). It is apparent that the current facilities cannot meet the ATRCR requirement.

It can be seen from Table 1 that eight land use types exist in this study area. The runoff coefficient of each land use type was obtained based on field surveys and experience. According to Equation (1), the entire 9,987 m$^3$ of runoff needs to be controlled in order to obtain the 26 mm rainfall depth. Therefore, there remains 8,258 m$^3$ of runoff that needs to be controlled by additional LID facilities.

**Table 1**

Amount of runoff from different land use types.

| No. | Land use type | Runoff coefficient | Surface area (ha) | Comprehensive runoff coefficient | Amount of runoff (m³) |
|---|---|---|---|---|---|
| 1 | Asphalt roof | 0.90 | 12.56 | | 2,938 |
| 2 | Concrete or asphalt pavement | 0.90 | 22.38 | | 5,236 |
| 3 | Courts and track field | 0.90 | 2.76 | | 647 |
| 4 | Dry masonry pavement | 0.40 | 0.77 | | 81 |
| 5 | Parking lot | 0.20 | 3.15 | | 164 |
| 6 | Training field | 0.25 | 0.96 | | 62 |
| 7 | Football field | 0.15 | 1.43 | | 55 |
| 8 | Greenland | 0.15 | 20.59 | | 803 |
| **Total** | | | 64.61 | 0.59 | 9,987 |

Considering the goals of the sponge city and site conditions, i.e., hydrology, land use, historical inundation points, drainage system, and space availability (Eckart et al., 2017; USEPA, 2000) as mentioned in the introduction and section 2.1, five LID facilities were selected, including bio-retention facilities, grassed swales, sunken green spaces, permeable pavements, and storage tanks. These LID facilities with various locations and sizes constituted five scenarios that examined how the variation of LID facilities in the catchment would affect the LID comprehensive performance and thus provided insights into ways to optimize design schemes (Fig. 4). In particular, the bio-retention facility and sunken green space that can not only mitigate peak flows effectively but also purify rainwater and enhance the landscape were located in the northwest corner of the sports center where inundation frequently occurred to retain rainfall runoff (DeBusk and Wynn, 2011; He et al., 2013). Also, there is no existing green land but large collecting area around the main stadium, making the runoff volume considerable during wet seasons and therefore the grey facilities of storage tanks, with less space demand and large retention and storage capacities, were located around the stadium (Damodaram et al., 2010). Additionally, grassed swales that can reduce and convey runoff and permeable pavements allowing rainfall infiltration were also selected (Davis et al., 2012a; Fassman and Blackbourn, 2010; Imran et al., 2013).

Insert Fig. 4

**Fig. 4.** Illustration of the distribution of LID facilities for each scenario.

In this case, as shown in Table 2, scenario 1 and 2 were designed as green-grey strategies where the sizes of LID facilities are at medium levels among five scenarios, whereas scenario 3 was designed representing a grey strategy that incorporates large areas of storage tanks. In contrast, scenario 4 and 5 were designed as a green strategy with large areas of bio-retention facilities and sunken green spaces, and grassed swales, respectively. These scenarios were designed and simulated by the size of the area affected by LID facilities. In order to determine the sizes of LID facilities in each scenario, the runoff control capacity of each LID facility per unit area was first distributed according to engineering practical experience and LID parameters (e.g., storage thickness) referring to SWMM user's manual by the US EPA (USEPA, 2015) and Sponge City technical manual (MOHURD, 2014). Details of the model parameters are provided in the supplementary material (Section 2). The bio-retention facility per unit area in $m^2$ can control 0.3 $m^3$ of runoff. The grassed swale, the sunken green space, and the permeable pavement per unit area in $m^2$ have the runoff control capacities of 0.15 $m^3$, 0.25 $m^3$, and 0.05 $m^3$, respectively. Also, the storage tank per unit area in $m^2$ can store 1 $m^3$ of runoff assuming its depth is 1 m. Therefore, the 8,258 $m^3$ of runoff were allocated to these five facilities in each scenario and thus the surface area of each LID facility was allocated (Table 2).

**Table 2**

Allocated surface area of each LID facility in each scenario.

| Scenario | Bio-retention facility (ha) | Grassed swale (ha) | Sunken green space (ha) | Permeable pavement (ha) | Storage tank (ha) | Total area (ha) |
|---|---|---|---|---|---|---|
| 1 | 0.757 | 0.429 | 0.847 | 0.142 | 0.316 | 2.490 |
| 2 | 1.035 | 0.529 | 0.941 | 0.243 | 0.189 | 2.936 |
| 3 | 0.850 | 0.350 | 0.305 | 0.325 | 0.426 | 2.256 |
| 4 | 1.125 | 0.480 | 1.500 | 0.120 | 0.035 | 3.261 |
| 5 | 0.675 | 0.780 | 0.850 | 0.248 | 0.281 | 2.835 |

### 2.3. Evaluation of individual benefits

Before the evaluation of the comprehensive performance of the LID scenarios, an individual benefit evaluation system was developed that involved environmental, economic, and social benefits.

### 2.3.1. Selection of evaluation indicators

A hierarchical structure of evaluation indicators was formed, as shown in Fig. 5. The hierarchy consisted of several levels of indicators and corresponding sub-indicators. The rationale for the selection of these indicators is discussed below.

**Environmental benefits**

LID facilities can effectively control the quantity and quality of rainfall runoff (Bedan and Clausen, 2009; Demuzere et al., 2014; Xu et al., 2017). For water quantity, three indicators were selected because they are the main factors that affect runoff control efficiencies: the runoff reduction, peak reduction, and peak delay (Bosley II, 2008; Qin et al., 2013). For water quality, the typical pollutants in rainfall runoff that affect the water quality of receiving waters are total suspended solids (TSS), chemical oxygen demand (COD), total nitrogen (TN), and total phosphorus (TP) (Mao et al., 2017). Hence, these were chosen as the relevant indicators (Fig. 5).

**Economic benefits**

The application of LID facilities can reduce the investment that municipal networks and facilities have to spend for the conveyance and treatment of stormwater (Eckart et al., 2017; Roy et al., 2008; Wang et al., 2013). Additionally, these design components can reduce maintenance costs in comparison with traditional stormwater control projects (DoD, 2015). The economic benefits of LID facilities are also reflected in the operation performance (Åstebøl et al., 2004), which is mainly determined by design feasibility, engineering feasibility, and the operational stability of LID facilities. Therefore, the economic benefits primarily include reductions in construction costs, maintenance costs, and savings due to increased operational performance (Fig. 5).

**Social benefits**

Since LID facilities have the capacity of storing rainwater which can be utilized, their application will therefore alleviate some of the water scarcity problems (He et al., 2013; Li et al., 2010). Also, LID facilities such as bio-retention facilities and grassed swales, offer attractive green views compared to conventional grey infrastructure (ESD, 2007). Hence, they enhance the value of the landscape. Moreover, LID facilities provide suitable conditions for plant growth and microorganism reproduction, which aids to ensure the health of the ecological environment (Ignatieva et al., 2011). Also, the facilities promote a more harmonious relationship between humans and nature (Benedict and McMahon, 2006). Therefore, the social benefits of LID facilities consist of water reuse functions, landscape functions, and ecological service functions (Fig. 5).

Insert Fig. 5

**Fig. 5.** The framework of the hierarchical structure of indicators.

**2.3.2. Determination of indicator weights**

Selected indicators were weighted using the AHP method, which is one of the multiple-criteria decision analysis methods that developed by Saaty (2008), evaluating multiple criteria explicitly for structuring and solving decision problems using a scoring and sorting process. In this case, original scores of the indicators were obtained using a questionnaire given to experts in the relevant field based on local water environment issues, the rainwater reuse demand, and the characteristics of LID facilities such as cost and operation performance (Chen et al., 2015; Phua and Minowa, 2005). The obtained scores were then used to sort indicators by calculating their weights using pairwise comparison matrix operations. A detailed description of the AHP method can be found in the literature (Linkov and Moberg, 2011; Rahmati et al., 2016). To ensure the reliability of the AHP method, the Consistency Ratio (CR) was calculated to examine the consistent extent of the expert judgment on the indicator scores (Ishizaka and Siraj, 2018; Jaiswal et al., 2014). If the CR is less than 0.1, the matrix is considered reliable, and the obtained weights can be used for further evaluation (Rahmati et al., 2016). Details of the AHP method are provided in the supplementary material (Section 1).

**2.3.3. Determination of indicator values**

The individual benefits of the LID scenarios were evaluated by quantifying the indicators mentioned above. The indicator values of environmental benefits were obtained using the SWMM, while indicator values of economic and social benefits were obtained according to the size of the LID facilities in each scenario (Table 2) and the weights of the LID facilities in view of each indicator derived from the AHP method, as discussed in Section 2.3.2.

To evaluate the environmental benefits, the performance of water quantity and quality control was simulated for each LID scenario. In the supplementary material (Section 1), according to local hydrological conditions and the distribution of the existing stormwater pipe networks, the study area was roughly divided into 84 sub-catchments. Under the condition of 75% ATRCR, a 26 mm rainfall event was used as one of the inputs for the model to examine the potential capacity of LID facilities regarding controlling annual total runoff from a statistical point of view. Also, rainfall events that decrement and increment by 10 mm, i.e., 16 mm and 36 mm rainfall events, were also used to evaluate the performance of the LID facilities under lower and upper boundary rainfall conditions. The hyetograph of rainfall events was designed using the Chicago Approach, which is commonly adopted in China (Xie et al., 2017). Rainfall duration was set to 90 minutes to simulate extreme rainfall events with high intensity, while the peak timing ratio was set to 0.5 according to actual rainfall patterns in the study area. If the LID practices could control the design rainfall event of 26 mm, it somehow indicates that the LID practices can control 26 mm of rainfall in each rainfall event (or the total rainfall if a rainfall event is less than 26 mm) throughout the year and thus corresponds to a 75% annual total runoff control rate. Fig. 6 shows the hyetograph of the three designed rainfall events.

Insert Fig. 6

**Fig. 6.** Hyetograph of the designed rainfall events.

In SWMM, model parameters were selected according to local-specific conditions (e.g., stormwater, land-use, and soil characteristics), with reference to SWMM user's manual by the US EPA (USEPA, 2015) and Sponge City technical manual (MOHURD, 2014). Therefore, the Horton infiltration model was used to determine infiltration and hence the runoff volume, while the SWMM routing model was selected to simulate the runoff routing process (He et al., 2013). In addition to the runoff model, the surface pollutant model was used to simulate the buildup and wash-off of pollutants, including TSS, COD, TN, and TP, on the urban catchment surfaces of roads, roofs, and green land. The saturation function model was applied to simulate the pollutant buildup process, while the exponential function model was used to represent the wash-off process (Baek et al., 2015).

Since the SWMM was used to conduct a comparative analysis for LID scenarios rather than real situations, the model parameters could not be calibrated using monitoring data. Instead, the basic model parameters were validated graphically and statistically using an observed rainfall event before application (Baek et al., 2015). During the rainfall event, indicators of flow rate (L/s), TSS (mg/L), COD (mg/L), TN (mg/L), and TP (mg/L) were measured in manholes at the outfall of A,

C, and E (Fig. 2). Results of the validation indicated that the Nash-Sutcliffe Efficiency (NSE) is greater than 0.5, and thus the applied model is considered acceptable (Moriasi et al., 2007). Details of the model validation are provided in the supplementary material (Section 3).

### 2.3.4. Calculation of individual benefits

The individual benefits were calculated using a weighted summation of indicator values, which is given by Equation (2):

$$B_{im} = \sum_k W_{ki} \cdot I_{km} \qquad (2)$$

where $B_{im}$ is the benefit $i$ in scenario m, $W_{ki}$ is the calculated weight of indicator $k$ in benefit $i$, and $I_{km}$ is the normalized value of indicator $k$ in scenario $m$ calculated using the method of linear normalization to make the indicator values dimensionless, which is given by Equation (3):

$$I_{km} = \frac{X_{km}}{\sum_{m=1}^{M} X_{km}} \qquad (3)$$

where $X_{km}$ is the original value of the indicator $k$ in scenario $m$, and $M$ is the number of the LID scenario (i.e., $M=5$ in this case).

### 2.4. Evaluation of comprehensive benefits

The comprehensive benefit evaluation depended on the evaluation of individual benefits, and the evaluation procedure is summarized in Fig. 7. First, sponge city goals were established before investigating the site conditions: hydrology, the types of land use, historical inundation points, and the distribution of drainage systems. Then LID scenarios were developed based on the basic requirements of the ATRCR. The evaluation system was developed based on the environmental, economic, and social benefits of LID practices. Following this, benefit indicators were selected and weighted. Then indicator values for each LID scenario were calculated. The individual and comprehensive benefits of each scenario were evaluated successively based on Equation (2) introduced in Section 2.3.4. Evaluation results were validated to ensure that the optimal LID scheme could meet the overall goals of a sponge city rather than only meet the ATRCR requirement. If these goals were not met, then the LID scenarios would need to be redeveloped.

Insert Fig. 7

**Fig. 7.** The procedure of the comprehensive performance evaluation.

## 3. Results and discussion

### 3.1. Indicator weights

The comprehensive benefit is composed of environmental, economic, and societal benefits, and the weights of their indicators and corresponding sub-indicators were calculated and are shown in Table 3. At each step in the hierarchy, the sum of the indicator weights should be one. The higher the value of the weight is, the more important the indicator is.

Considering the goals of sponge city construction (Ahiablame et al., 2013; Xie et al., 2017), the indicator of environmental benefit had the highest weight (0.608) among its counterparts. Also, due to the existing inundation issues, the sub-indicator of water quantity had a weight of 0.700, which was much greater than the 0.300 weight of water quality. Moreover, sub-indicators of runoff reduction, peak reduction, and peak delay had weights of 0.607, 0.303, and 0.090, respectively. In contrast, the indicator of economic benefit ranked second (0.272). Since the feasibility of design and construction and the operation stability are essential indicators that reflect the cost-effectiveness of a given LID facility (Hirschman and Woodworth, 2010), the sub-indicator of operation performance had the highest weight (0.655). The indicator of social benefit had a relatively small weight (0.120), and its corresponding sub-indicator of water reuse function had the highest weight (0.648) because of water reuse demand in the study area.

**Table 3**

The weighting system for the comprehensive evaluation of the sponge city.

| | Indicators | Weights | Sub-indicators | Weights | Sub-indicators | Weights |
|---|---|---|---|---|---|---|
| | | | | | Runoff reduction | 0.607 |
| | | | Water quantity | 0.700 | Peak reduction | 0.303 |
| | | | | | Peak delay | 0.090 |
| | Environmental benefit | 0.608 | | | TSS reduction | 0.466 |
| | | | Water quality | 0.300 | COD reduction | 0.277 |
| | | | | | TN reduction | 0.161 |
| | | | | | TP reduction | 0.096 |
| Comprehensive benefit | | | Construction cost | 0.187 | | 0.187 |
| | | | Maintenance cost | 0.158 | | 0.158 |
| | Economic benefit | 0.272 | | | Design feasibility | 0.200 |
| | | | Operation performance | 0.655 | Engineering feasibility | 0.400 |
| | | | | | Operation stability | 0.400 |
| | | | Water reuse function | 0.648 | | 0.648 |
| | Social benefit | 0.120 | Landscape function | 0.122 | | 0.122 |
| | | | Ecological function | 0.230 | | 0.230 |

The CR of the matrix that derived the comprehensive benefits was 0.03, which is smaller than 0.1, indicating that the matrix is reliable to obtain weights. Also, the CR of the matrix, which derives the sub-indicators of environmental, economic, and social benefits, was 0, 0.07, and 0.07, respectively. The CR of the matrix that derived the sub-indicators of water quantity, water quality, and operation performance was 0.03, 0.01, and 0.06, respectively. Overall, the CRs of these matrixes were all smaller than 0.1, and hence the assigned weights based on the AHP method are reliable.

## 3.2. Individual benefits of different LID scenarios

### 3.2.1. Environmental benefits

The environmental benefits represented by the indicators of water quantity and quality were simulated to compare their rate of change before and after the addition of LID facilities in the study area (Table 4). The results presented in Table 4 were derived from the six outfalls (Fig. 2) and obtained by averaging the original results of three rainfall events considering that observed rainfall events can be less or greater than the design rainfall event of 26 mm. Scenario 4 shows the best performance among its counterparts in terms of reducing total runoff ($m^3$), peak flow (L/s), and all pollutants (mg/L). Scenario 4 mainly consisted of two LID facilities, i.e., a bio-retention facility and a sunken green space, where the bio-retention facility and the sunken green space together accounted for 80.5% of the total surface area of the LID facilities. This result is in accordance with the literature that suggests that a bio-retention facility coupled with a permeable pavement can reduce the total runoff and peak flow by 40% (W. Li et al., 2017). Moreover, scenario 1 ranked second of all the indicators except peak time delay, and the sunken green space, which accounted for 34.0% of the total LID area, can slow down the conveyance of collected runoff to infiltrate the soil below. This percentage was higher than that for scenarios 2, 3, and 5, whereas this was not the case for the bio-retention facility. This was likely due to the runoff control efficiency of the sunken green space being greater than that of the bio-retention facility, which is in accordance with the literature (Lin et al., 2016). The peak delay time of scenario 1 was the greatest, although the surface areas of the bio-retention facility and the sunken green space were less than those in scenarios 2 and 4. This was mainly due to its large 3160 $m^3$ storage tanks that were designed to collect runoff during rainfall events and to drain the rainwater during dry periods, and the location of the storage tanks that were evenly distributed around the stadium and thus lead to different control efficiencies of rainfall runoff from the stadium (Xing et al., 2016). Scenario 3 had the largest areas of permeable pavements and storage tanks, while its performance is similar to that of scenario 5 which had the largest areas of grassed swales. This is probably due to characteristic parameters of LID practices that allow the permeable pavements together with storage tanks could reduce flow and pollutants through infiltration, retention, and under-drainage (Nichols et al., 2015), which showed a comparable capability with grassed swales that could reduce flow and pollutants through interception, infiltration, storage, and conveyance (Davis et al., 2012a).

**Table 4**

Simulation-based indicator values of the environmental benefit for each scenario.

| Scenario | Runoff reduction | Peak flow reduction | Peak time delay | TSS reduction | COD reduction | TN reduction | TP reduction |
|---|---|---|---|---|---|---|---|
| 1 | 19.3% | 20.9% | 3 min | 21.6% | 22.3% | 20.5% | 18.7% |
| 2 | 17.1% | 18.7% | 1 min | 19.7% | 20.2% | 18.5% | 16.5% |
| 3 | 17.8% | 18.3% | 1 min | 19.7% | 19.2% | 18.6% | 16.9% |
| 4 | 19.6% | 21.8% | 2 min | 22.4% | 23.1% | 21.3% | 19.2% |
| 5 | 17.3% | 18.5% | 1 min | 19.3% | 19.9% | 18.2% | 16.4% |

### 3.2.2. Economic and social benefits

Indicator values of the economic and social benefits of each scenario were obtained according to the sizes and weights of the LID facilities and then were normalized using Equation (3), as shown in Table 5.

It can be seen that scenario 3 had the highest engineering feasibility of 0.211 among these scenarios, which is primarily due to the application of large 4260 m$^3$ storage tanks needing simple engineering process compared to others such as the bio-retention facility with multilayer structures, although it is relatively not cost-effective. In contrast, scenario 4 had the highest economic benefit for a design feasibility of 0.216 and an operation stability of 0.231, as well as social benefits. In addition, the landscape function was almost twice as large as that of scenario 3, mainly because scenario 4 had large areas of bio-retention facilities which show good operation performance in many cases (DeBusk and Wynn, 2011; Liu et al., 2014). The landscape function of a bio-retention facility is respectable and differs from conventional urban gardens due to its attractive appearance and diversification of plant species (Kim and An, 2017).

**Table 5**

Normalized indicator values of the economic and social benefits for each scenario.

| Scenario | Construction cost | Maintenance cost | Design feasibility | Engineering feasibility | Operation stability | Water reuse function | Landscape function | Eco−service function |
|---|---|---|---|---|---|---|---|---|
| 1 | 0.202 | 0.174 | 0.181 | 0.186 | 0.181 | 0.188 | 0.177 | 0.176 |
| 2 | 0.199 | 0.222 | 0.215 | 0.207 | 0.215 | 0.211 | 0.221 | 0.220 |
| 3 | 0.226 | 0.185 | 0.200 | 0.211 | 0.185 | 0.204 | 0.154 | 0.158 |
| 4 | 0.182 | 0.236 | 0.216 | 0.191 | 0.231 | 0.212 | 0.268 | 0.257 |
| 5 | 0.192 | 0.182 | 0.188 | 0.206 | 0.189 | 0.185 | 0.179 | 0.189 |

### 3.3. Comprehensive benefits of different LID scenarios

Individual and comprehensive benefits of each scenario were calculated using a weighted summation of the normalized indicator values successively, which are shown in Table 6. While the numerical difference among these scenarios was small, their difference is significant regarding the practical application (Ahiablame et al., 2013).

It can be seen that scenario 4 had the maximum comprehensive benefit of 0.218, the largest social benefit, and a relatively large environmental benefit when compared with scenarios 2, 3, and 5. Scenario 4 also had an economic benefit that was larger than other scenarios except scenario 2. Scenario 1 came next, with a comprehensive benefit of 0.208. Scenario 1 also had the highest environmental benefit, whereas it showed poor economic and social benefits among its counterparts which is mainly attributed to its small areas of LID practices. In contrast, scenario 5 had the minimal comprehensive benefit and environmental benefit, due to the small areas of the bio-retention facility and the sunken green space. These areas accounted for only 60% and 57% of the corresponding facilities in scenario 4. Scenario 2 had the highest economic benefit, mainly because it consisted of the second largest area of the bio-retention facility and the sunken green space, and relatively larger grassed swales than that in scenario 4. This illustrated the contributions of the bio-retention facility and the sunken green space to good comprehensive performance, which has also been suggested in the literature (He et al., 2013; J. Li et al., 2017).

**Table 6**

Individual and comprehensive benefits of each scenario.

| Scenario | Environmental benefit | Economic benefit | Social benefit | Comprehensive benefit |
|---|---|---|---|---|
| 1 | 0.222 | 0.185 | 0.184 | 0.208 |
| 2 | 0.186 | 0.212 | 0.215 | 0.196 |
| 3 | 0.187 | 0.201 | 0.188 | 0.191 |
| 4 | 0.220 | 0.210 | 0.229 | 0.218 |
| 5 | 0.185 | 0.192 | 0.185 | 0.187 |

Overall, in terms of providing maximum comprehensive benefits for a sponge city, scenario 4 was considered the optimal LID scheme to be applied to the sports center project, with the bio-retention facility accounts for 34.5% of the total surface area of the LID facilities and the sunken green space accounts for 46.0% of the total area. Since scenario 1 had a smaller comprehensive benefit and, at the same time, a smaller total area than that of scenario 4, the results suggested that scenario 1 could be an alternative. According to the ratio of the area of each LID facility to the total LID facilities in scenario 1 and 4, the results suggested that the surface area proportion of the bio-retention facility can range from 30% to 35%, while the proportion for the sunken green space can range from 34% to 46%. These ranges can inform the optimization of LID facilities in other sponge city projects.

On the basis of the proposed evaluation system, the construction of a sponge city can achieve respectable comprehensive performance. However, in order to apply the LID scheme effectively, more detailed modeling and field studies should be considered before practical uses of the results. In terms of model simulation, the design rainfall events used in this case may lead to limitations in evaluating the runoff control performance of LID scenarios. Therefore, long-term and high resolution (e.g., 1-min interval) rainfall monitoring data are preferable (Mao et al., 2017). Also, limitations of models should also be taken into account when evaluating the simulated results. In this case, the slight difference in peak time delay among different scenarios (Table 4) could result from the model limitation in simulating peak flows (Barco et al., 2008). This may also be, however, a challenge for other models when simulating hydraulic and hydrological mechanism in urban areas due to large impervious surfaces (Ackerman et al., 2005). More consideration should be therefore given to model selection and validation.

## 4. Conclusions

The construction of a sponge city can contribute to the overall comprehensive benefits in the process of developing a sustainable city. A proposed comprehensive evaluation system was implemented in the study area that quantified the performance of various LID combination scenarios with respect to the environmental, economic, and social benefits. Results indicated that the green scenario (4) that contains 34.5% of bio-retention facilities and 46.0% of sunken green spaces had the best comprehensive benefit as well as social benefit. In contrast, the green scenario (5) had large areas of grassed swales but the lowest comprehensive benefit, mainly because the bio-retention facilities and sunken green spaces accounted for only 60% and 57% of the corresponding facilities in scenario 4. In terms of providing the environmental and economic benefits, the green-grey scenarios (1 and 2) showed better performance than green scenarios (4 and 5). Although the grey scenario (3) spent minimal spaces, its performance in providing both the individual and comprehensive benefits was mediocre. The optimal scheme (scenario 4) can not only meet the rainfall runoff control requirement but also provide respectable economic and social benefits including good operation performance, rainwater utilization, landscape promotion, and ecological service functions, mainly because the two main facilities can manage stormwater at the source through the natural processes of infiltration, detention, storage, and purification. Based on the top two scenarios regarding comprehensive benefits, proportional ranges of the area of the bio-retention facility and the sunken green space to total LID facilities were also suggested, which should be in the range of 30% to 35% and 34% to 46%, respectively. These ranges are a reference for the optimization of LID facilities in other sponge city projects.

The optimal LID scenario obtained from the proposed evaluation system can be specific to each region due to local-specific project goals and site conditions. However, the evaluation system can provide insights into ways to take into account the comprehensive benefits of LID practices and to optimize design schemes before sponge city construction in other regions. Moreover, it can also be available for regions where sponge city projects have been built, and the performance effect evaluation of LID practices is needed. Future research will concentrate on long-term monitoring of rainfall data for robust model simulation, validation, and calibration in order to better evaluate the performance of LID practices for sponge city construction.

## Appendix A. Supplementary data
Supplementary data related to this manuscript can be found in the file "Supplementary Material."

## Acknowledgments
This study was financially supported by National Water Grant "Major Science and Technology Program for Water Pollution Control and Treatment" (No. 2017ZX07202002) and the Shenzhen Science and Innovation Commission (JSGG20170412145935322 and JSGG20160428181710653). Special thanks to the Development and Reform Commission of Shenzhen Municipality (Urban Water Recycling and Environment Safety Program) and the Transport Commission of Shenzhen Municipality.